\documentclass{article}
\usepackage[english]{babel}
\usepackage{amsfonts}
\usepackage{amssymb}
\usepackage[dvips]{graphicx}
\usepackage{subfigure}
\usepackage{color}

\begin{document}

\title{Fluctuations and response in a non-equilibrium micron-sized system}
\author{Juan Ruben Gomez-Solano$^{1}$, Artyom Petrosyan$^{1}$, Sergio
Ciliberto$^{1}$ and Christian Maes$^{2}$ \\
Email: juan.gomez\textunderscore solano@ens-lyon.fr, artyom.petrosyan@ens-lyon.fr,\\
sergio.ciliberto@ens-lyon.fr and christian.maes@fys.kuleuven.be\\
$^1$Laboratoire de Physique, Ecole
Normale Sup\'erieure de  Lyon, CNRS UMR 5672,\\
        46, All\'ee d'Italie, 69364 Lyon CEDEX 07, France\\
$^2$Instituut voor Theoretische Fysica, K. U. Leuven, B-3001 Leuven, Belgium}
\maketitle

\begin{abstract}

The linear response of non-equilibrium systems with Markovian dynamics satisfies a generalized fluctuation-dissipation
relation derived from time symmetry and antisymmetry properties of the fluctuations. The relation
involves the sum of two correlation functions of the observable of interest: one with the entropy excess and the second
with the excess of dynamical activity with respect to the unperturbed process, without recourse to anything but the dynamics of the system. We illustrate this approach in the experimental determination of the linear response of the potential energy of a Brownian particle
in a toroidal optical trap. The overdamped particle motion is effectively confined to a circle, undergoing a periodic potential
and driven out of equilibrium by a non-conservative force. 
Independent direct and indirect measurements of the linear response around a non-equilibrium steady state are performed 
in this simple experimental system.
The same ideas are applicable to the measurement of the response of more general non-equilibrium micron-sized
systems immersed in Newtonian fluids either in stationary or non-stationary states and possibly including inertial
degrees of freedom.
\end{abstract}

\section{Introduction}

The linear response of systems in thermodynamic equilibrium is
generally described by the fluctuation-dissipation theorem \cite{callen}.
It provides a simple relation between the equilibrium
fluctuations of an observable $Q$ with the response 
due to a small external perturbation $h_s$ changing the
potential at time $s$ as $U \rightarrow U - h_s V$:
\begin{equation}\label{eq:FDT}
    R_{QV}(t-s)= \beta \frac{\mathrm{d}}{\mathrm{d}s} \langle Q(t) V(s) \rangle_{0}.
\end{equation}
In equation~(\ref{eq:FDT}) $R_{QV}(t-s)=\delta \langle Q(t) \rangle_h /
\delta h_s |_{h=0}$ is the linear response function of $Q$ at time $t \ge s$; 
$\langle Q(t) V(s) \rangle_{0}$ is the two-time
correlation function between $Q$ and $V$ measured at equilibrium;
the brackets $\langle \ldots \rangle_h$ denote the ensemble
average in the state perturbed by $h_s$ so that $\langle \ldots
\rangle_0$ corresponds to the ensemble average at equilibrium
($h_s = 0$). The inverse temperature of the equilibrium system,
$\beta = 1/k_B T$, appears as a multiplicative factor. Hence,
equation~(\ref{eq:FDT}) represents a useful tool in experiments and
simulations to explore indirectly the linear response regime from
fluctuation measurement completely performed at thermal
equilibrium. Vice versa, one can obtain information on microscopic
fluctuations from non-equilibrium measurements of
response functions or susceptibilities by applying sufficiently
weak external fields.

In general, equation~(\ref{eq:FDT}) fails to describe the linear
response of systems already prepared in a non-equilibrium state.
This situation is relevant in real mesoscopic systems that usually
operate far from equilibrium due to either
non-conservative/time-dependent forces exerted by the experimental
apparatus or external flows and gradients applied at the
boundaries. For instance, the developement of micro and nano
techniques (\emph{e.g.} optical tweezers and atomic force
microscopes) has allowed one to mechanically manipulate colloidal
particles, living cells and single molecules of biological
interest with forces ranging from pN to fN. In this kind of
experiments, thermal fluctuations can be comparable or larger than
the typical external perturbations necessary to determine
$R_{QV}$. Then, for these systems it is more reliable in practice
to measure  non-equilibrium fluctuations than linear response
functions.

On the theoretical side, several works have recently dealt with the problem 
of the extension of the fluctuation-dissipation theorem around non-equilibrium 
steady states \cite{agarwal,harada,lippiello,seifert1,chetrite,martens,puglisi1,villamaina,prost,seifert2}. 
In most of the formulations, an additive extra term on the right-hand side of equation~(\ref{eq:FDT}) 
appears as a non-equilibrium correction of the fluctuation-dissipation relation due to the broken 
detailed balance. Different physical interpretations of the corrective term are provided in the 
literature. We specially highlight
the roles of the non-vanishing probability current \cite{chetrite}
and the conjugate variables to the total entropy
production \cite{seifert1,seifert2} in the analytical expression
of the corrective term, leading independently to a Lagrangian
interpretation for Langevin systems with first-order Markovian dynamics.
Following these two approaches an equilibrium-like
fluctuation-dissipation relation can be restored in the Lagrangian
frame of the local mean velocity of the system for the right
choice of observables \cite{seifert1,chetrite,chetrite2}. An alternative
interpretation, the \emph{entropic-frenetic} approach, has been proposed in terms of time-symmetric and
time-antisymmetric properties of the fluctuations
\cite{baiesi1,baiesi2,baiesi3}. 
Unlike previous formulations, the entropic-frenetic approach does not involve the 
stationary probability distribution of the degree of freedom of interest but only explicit observables.
In addition, it holds even in more general situations including non-steady states and inertial
degrees of freedom provided that the dynamics is Markovian
\cite{baiesi3}. Therefore it is more accessible from the experimental point of view.

The present paper provides the first experimental investigation of the modified fluctuation-response relation of \cite{baiesi1,baiesi2}  in a simple experimental system driven into a non-equilibrium steady state: a Brownian particle whose overdamped motion is effectively confined to a circle by a toroidal optical trap \cite{gomez}.  As the new formula relating response and fluctuations is exact and does not involve further notions such as effective temperatures, it can be tested at full value.  Moreover,  we do not need the explicit knowledge (analytic form) of the stationary distribution.
Hence the experimental linear response of the system can be safely determined by two independent methods. In the first method one control parameter of the non-equilibrium steady state is physically perturbed and the corresponding response of the particle is directly measured. In the second method, only the unperturbed non-equilibrium fluctuations of the position of the particle are measured and the same linear response is determined based on the generalized fluctuation-dissipation formula and on a suitable model of the dynamics.
All of this is done here under very well controlled experimental conditions, so that all can be checked explicitly and every step can be analyzed separately.  That control allows us to understand the usefulness of the new formalism so that it will be able to be employed also in natural situations where such a control is really absent, and \emph{e.g.} the driving or potential are not exactly known.

The paper is organized as follows: in section~\ref{sec:generalized} we briefly present the generalized approach to linear response based on the entropic-frenetic formulation of a fluctuation-dissipation relation for systems with Markovian dynamics and the main goals of our experimental study with respect to this formalism. In section~\ref{sec:colloidal} we describe the main features of the experiment and the 1D Langevin model of the translational motion of the particle in the toroidal trap. In section~\ref{sec:linear} we present the results of the two methods to measure the linear response function of the system.  We show that within our experimental accuracy both methods lead to the same values of the linear response function when taking into account the corrections given by the generalized fluctuation-dissipation relation. Then, we discuss their technical limitations and advantages from the experimental point of view. We also depict a simple example in order to show the flexibility of the generalized fluctuation-dissipation formula: once the unperturbed fluctuations of the proper degree of freedom are measured, one can readily compute the linear response of the system under more complex time-dependent perturbations. Finally we present the conclusion.

\section{Generalized approach to linear response}\label{sec:generalized}
We briefly present the formulation of the fluctuation-dissipation
relation developed in \cite{baiesi1,baiesi2,baiesi3} for the
special case of stochastic systems described by a finite number of
degrees of freedom $\{q\}$ with overdamped Markovian Langevin
dynamics, in presence of a potential $U(q)$. We consider a system in
contact with a thermal bath at temperature $T$ and driven into a 
non-equilibrium steady state by a non-conservative force. We focus on
the average value of an observable $Q(q)$ at time $t$, denoted by
$\langle Q(q_t) \rangle_0$. We are also interested in the mean
value $\langle Q(q_t) \rangle_h$ of $Q(q)$, when a small time dependent
perturbation $h_s$ is applied to $U(q)$,
namely $U(q) \rightarrow U(q) - h_s V(q)$. As formally
shown in \cite{baiesi1,baiesi2}, $\langle Q(q_t) \rangle_h$ is
given at linear order in $h_s$ by
\begin{equation}\label{eq:linearresponse}
    \langle Q(q_t) \rangle_h = \langle Q(q_t) \rangle_0+\int_{-\infty}^t R_{QV}(t,s)h_s \, \mathrm{d}s,
\end{equation}
where the linear response function $R_{QV}$
obeys the generalized fluctuation-dissipation relation
\begin{equation}\label{eq:GFD}
    R_{QV}(t,s)=\frac{\beta}{2} \frac{\mathrm{d}}{\mathrm{d}s}\langle V(q_s)Q(q_t) \rangle_0 - \frac{\beta}{2} \langle LV(q_s) Q(q_t) \rangle_0.
\end{equation}
In equation~(\ref{eq:GFD}), $L$ is the generator of the unperturbed Langevin dynamics which determines the time evolution
of any single-time observable $O(q)$:  $\mathrm{d} \langle O(q_t) \rangle_0 / \mathrm{d}t = \langle (LO)(q_t) \rangle_0$.

It should be noted that $R_{QV}(t,s)$ is operationally obtained by applying an instantaneous delta
perturbation at time $s$ and measuring $\langle Q(q_t) \rangle_h - \langle Q(q_t) \rangle_0$.
In experiments it is always more reliable to apply a Heaviside perturbation
($h_s=0$ for $s<0$, $h_s=h=const.$ for $s\ge 0$) instead. This procedure directly yields the integrated
response function
\begin{equation}\label{eq:response}
    \chi_{QV}(t)=\int_0^t R_{QV}(t,s) \, \mathrm{d}s = \frac{\langle Q(q_t) \rangle_h-\langle Q(q_t) \rangle_0 }{h},
\end{equation}
defined over the time  interval $[0,t]$. Therefore, in the following we consider the integral form of equation~(\ref{eq:GFD})
\begin{equation}\label{eq:intGFD}
    \chi_{QV}(t)=\frac{\beta}{2} [ C(t) + K(t) ],
\end{equation}
where  the term
\begin{equation}\label{eq:entropic}
    C(t)=\langle V(q_t) Q(q_t)  \rangle_0 - \langle  V(q_0) Q(q_t) \rangle_0,
\end{equation}
can be interpreted as a correlation between the observable $Q(q_t)$ and the excess in entropy produced by the
Heaviside perturbation during the interval $[0,t]$: $[hV(q_t)-hV(q_0)]/T$. On the other hand, the term
\begin{equation}\label{eq:frenetic}
    K(t)=- \int_0^t \langle LV(q_s) Q(q_t) \rangle_0 \,\mathrm{d}s,
\end{equation}
can be interpreted as minus the correlation between 
$Q(q_t)$ and the integrated excess in dynamical activity or
\emph{frenesy}: $\beta\int_0^t LV(q_s)h \mathrm{d}s$, which
quantifies how frenetic the motion is due to the perturbation with
respect to the unperturbed process. The frenesy $\beta LV(q)$ can
be regarded as a generalized escape rate of a trajectory from a
given phase-space point $q$. In references
\cite{baiesi1,baiesi2,baiesi3} it has been widely discussed that the
origin of the entropic $C(t)$ and the frenetic $K(t)$ terms can be
traced back to time-antisymmetric and symmetric
properties of the fluctuations, respectively.

The non-equilibrium fluctuation-response formulae (\ref{eq:GFD}) and (\ref{eq:intGFD}) result directly from the theory of the Langevin equation, and hence, the validity of a Langevin model for our experiment would thus also imply this particular fluctuation-response formula.  Since for the present experiment the Langevin equation has been largely verified before, thus implying in theory the response  (\ref{eq:linearresponse})-(\ref{eq:frenetic}), the point of the present experimental study is not in the very first place to validate (\ref{eq:linearresponse})-(\ref{eq:frenetic}), although we still want to make sure it is exact. Rather, the point of the present investigation is  much more twofold:
\begin{itemize}
\item[1)] to show how the (very new) frenetic term (\ref{eq:frenetic}) can indeed be measured independently and directly, without recourse to anything but the dynamics.  In particular, no knowledge of the analytical expression of the stationary distribution is used;
\item[2)] to show how the new formalism (\ref{eq:linearresponse})-(\ref{eq:frenetic}) of response can be used to determine unknown forces, both conservative and non-conservative.  For example, when the formula is accepted as such, the correction to the equilibrium formula gives useful information about how the breaking of detailed balance gets established.  Alternatively, by changing the driving we can discover the potential that acts on the particle, etc.
\end{itemize}
These points are addressed in the following sections under proper experimental control and data analysis, which  allows us to understand how the generalized fluctuation-response formalism can be applied to a real non-equilibrium experiment.

\section{Colloidal particle in a toroidal optical trap}\label{sec:colloidal}

\begin{figure}
     \centering
     \subfigure[]{
          \includegraphics[width=.44\textwidth]{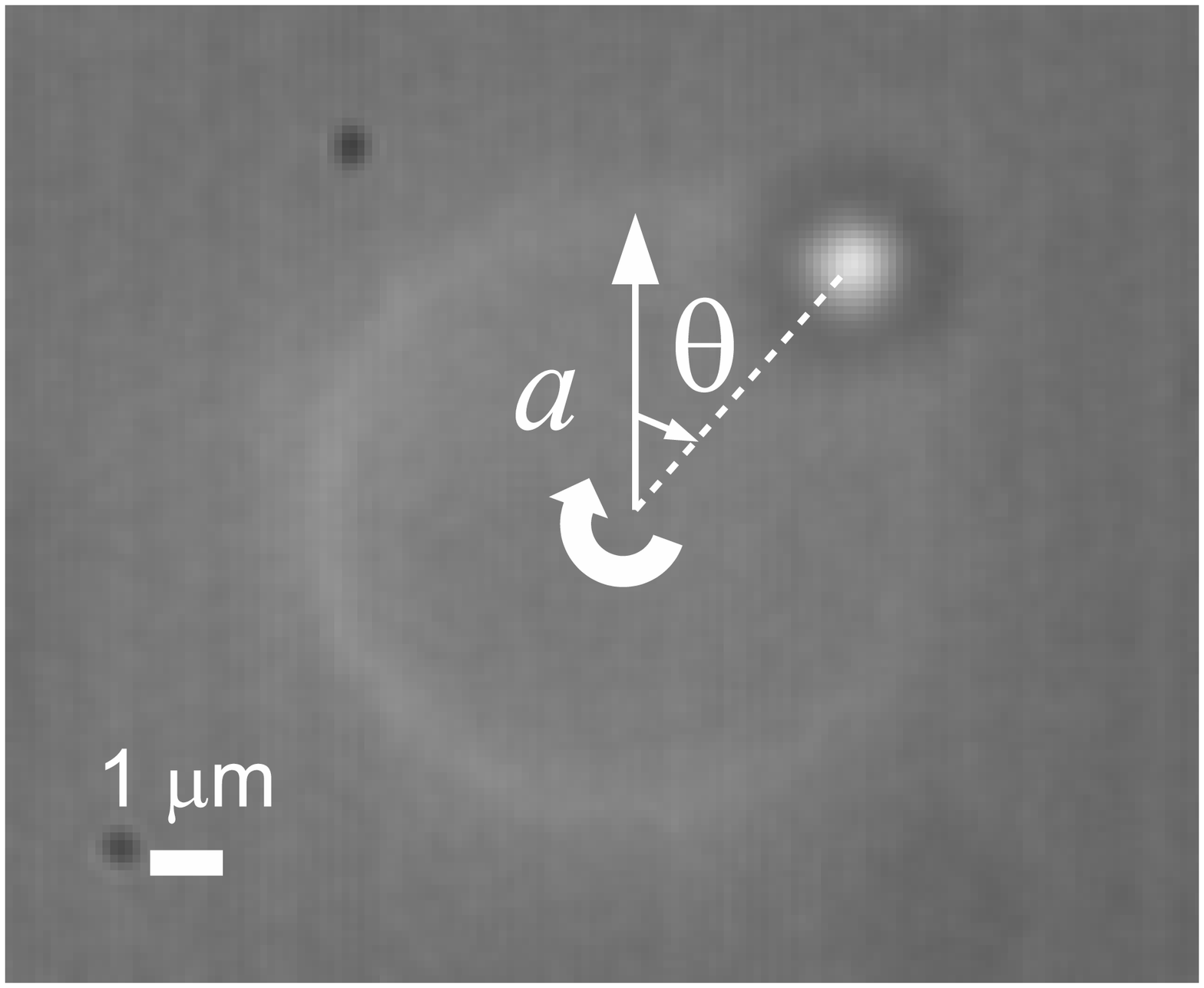}}
     \hspace{.0in}
     \subfigure[]{
          \includegraphics[width=.51\textwidth]{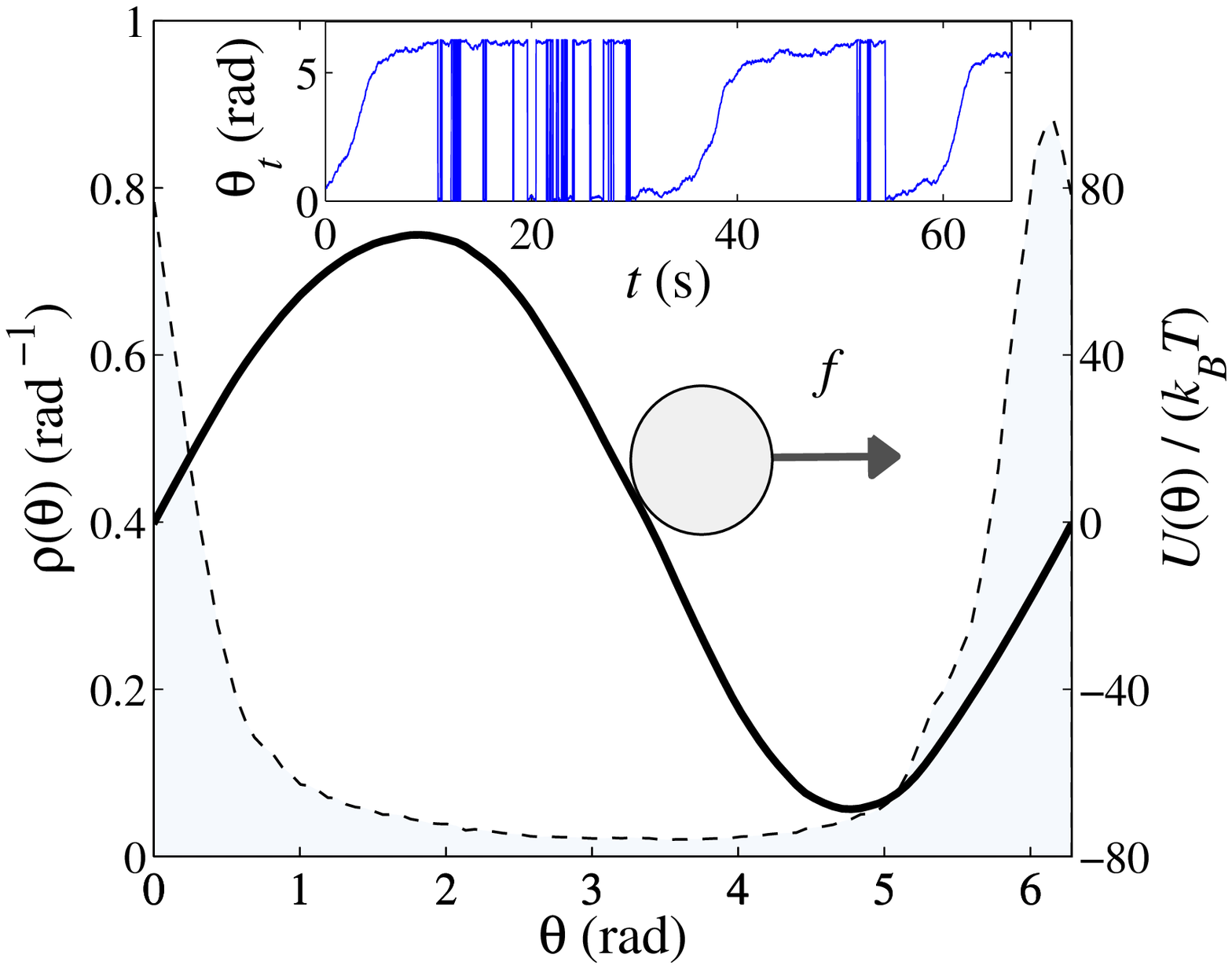}}\\
\vspace{.0in}
     \caption{ (a) Snapshot of the colloidal particle in the toroidal optical trap.
The vertical arrow indicates the position $\theta=0$ whereas the curled arrow shows the direction of the
rotation of the laser beam. (b) Experimental potential profile (solid line) and probability
density function of $\theta$ (dashed line) for the non-equilibrium steady state generated by the
non-conservative force $f$. Inset: Typical steady-state trajectory $\{\theta_t, \, 0\le t \le
66.67$ s$\}$ used to compute $\rho(\theta)$.}
     \label{fig:1}
\end{figure}

\subsection{Experimental description}
We specifically study the linear response of a single colloidal
particle driven out of equilibrium by a non-conservative force
in presence of a non-linear potential. We recall the main features of the experiment, previously
described in detail in reference \cite{gomez}, where the same experimental
set-up was used in the context of a modified fluctuation-dissipation relation with a different interpretation from the
one described in the present paper. In our experiment the
Brownian motion of a spherical silica particle (radius $r=1 \,
\mu$m) immersed in water is confined on a thin torus of major
radius $a=4.12 \, \mu$m by a tightly focused laser beam rotating at 200 Hz (see figure~\ref{fig:1}(a)). 
The rotation frequency of
the laser is so high that it is not able to trap continuously the
particle in the focus because the viscous drag force of
the surrounding water quickly exceeds the optical trapping
force. Consequently, at each rotation the beam only kicks the
particle a small distance along the circle of radius $a$. During
the absence of the beam ($\approx 5$ ms), the particle undergoes
free diffusion of less than 40 nm in the radial and perpendicular
direction to the circle. Thus, the particle motion is effectively
confined on a circle: the angular position $\theta$ of its
barycenter is the only relevant degree of freedom of the dynamics.
In addition, a static light intensity profile is created along the
circle by sinusoidally modulating in time the laser power which
has a mean value  of  30 mW and a modulation amplitude of 4.2 mWpp
at the same frequency as the rotation frequency of the beam.
Figure~\ref{fig:1}(a) sketches this experimental configuration on a
snapshot  of the colloidal particle in the toroidal trap. 
The water reservoir acts as a thermal bath at
fixed temperature ($T=20 \pm 0.5^{\circ}$C) providing thermal
fluctuations to the particle.
The viscous drag coefficient at this temperature is $\gamma=1.89 \times 10^{-8}$ kg s$^{-1}$.

\subsection{Model}
For the experimentally accessible length and time scales the dynamics of $\theta$ is modeled by
the first-order Langevin equation 
\begin{equation}\label{eq:1stLangevin}
     \dot{\theta} = -A \phi'(\theta)+F+ \xi,
\end{equation}
as extensively verified in \cite{gomez,blickle,blickle1,blickle2,mehl}. 
$A\phi(\theta)$ is a periodic non-linear potential
[$A\phi(\theta)=A\phi(\theta+2\pi)$] of amplitude $A = 0.87$
rad$^2$ s$^{-1}$ created by the periodic modulation of the laser intensity. 
The normalized angular profile $\phi(\theta)$ of the potential ($\max\{|\phi(\theta)|\}=1$)
is a slightly distorted sine function as a result of unavoidable experimental static defects of the
toroidal optical trap (\emph{e.g.} optical aberration).
$F = 0.85$ rad s$^{-1}$ is a constant force acting in the direction of
the laser rotation which is associated to the mean kick of the
beam. $\xi$ is a white noise process of zero mean and covariance
$\langle \xi_t \xi_s \rangle = 2D \delta(t-s)$ with bare
diffusivity $D=k_B T /(\gamma a^2)= 1.26 \times 10^{-2}$ rad$^2$
s$^{-1}$, which models the thermal fluctuating force exerted by
the water molecules. $F$ is non-conservative ($\int_0^{2\pi} F
d\theta = 2\pi F > 0$) since the motion takes place on a circle,
driving the system out of equilibrium. The physical
non-conservative force and the potential are $f=\gamma a F=66$ fN
and $U(\theta)=\gamma a^2 A \phi(\theta) = 68.8k_B T
\phi(\theta)$, respectively. The experimental potential profile
$U(\theta)$ is plotted as a continuous black line in
figure \ref{fig:1}(b). Note that at thermal equilibrium ($F=0$) the
particle motion would be tightly confined around the potential
minimum with the stochastic variable $\theta$ distributed
according to the Boltzmann density $\rho_{eq}(\theta) \propto
\exp[-\beta U(\theta)]$. However, due to the thermal
fluctuations and the non-conservative force $F>0$, the particle is
able to go beyond the potential barrier and explore the whole
circle. In the non-equilibrium situation with constant $F,A,D>0$,
the angular position $\theta$ settles in a stationary probability
density $\rho(\theta) \neq \rho_{eq}(\theta)$ that admits an
analytical expression found in \cite{maes}. The experimental
non-equilibrium density $\rho(\theta)$ is shown in
figure \ref{fig:1}(b) as a dashed line. A constant
probability current $j= \langle \dot{\theta}(t) \rangle_0/(2
\pi)=[F-A\phi'(\theta)]\rho(\theta)-D\partial_{\theta} \rho(\theta)]> 0$, 
in the direction of $F$ appears reflecting the broken
detailed balance of the dynamics. For the experimental conditions
one finds $j=3.76 \times 10^{-2}$ s$^{-1}$ corresponding to a mean
rotation period of 26.6 s for the particle. Both $\rho(\theta)$
and  $j$, determined from 200 independent experimental time series
of the angular position of the particle $\{\theta_t, \, 0\le t \le
66.67$ s$\}$, allow one to precisely compute the values of $f$ and $U(\theta)$,
as described in detail in \cite{gomez,blickle}.
One example of such a steady-state time series is depicted in the inset of figure \ref{fig:1}(b).

\begin{figure}
     \centering
     \subfigure[]{
          \includegraphics[width=.45\textwidth]{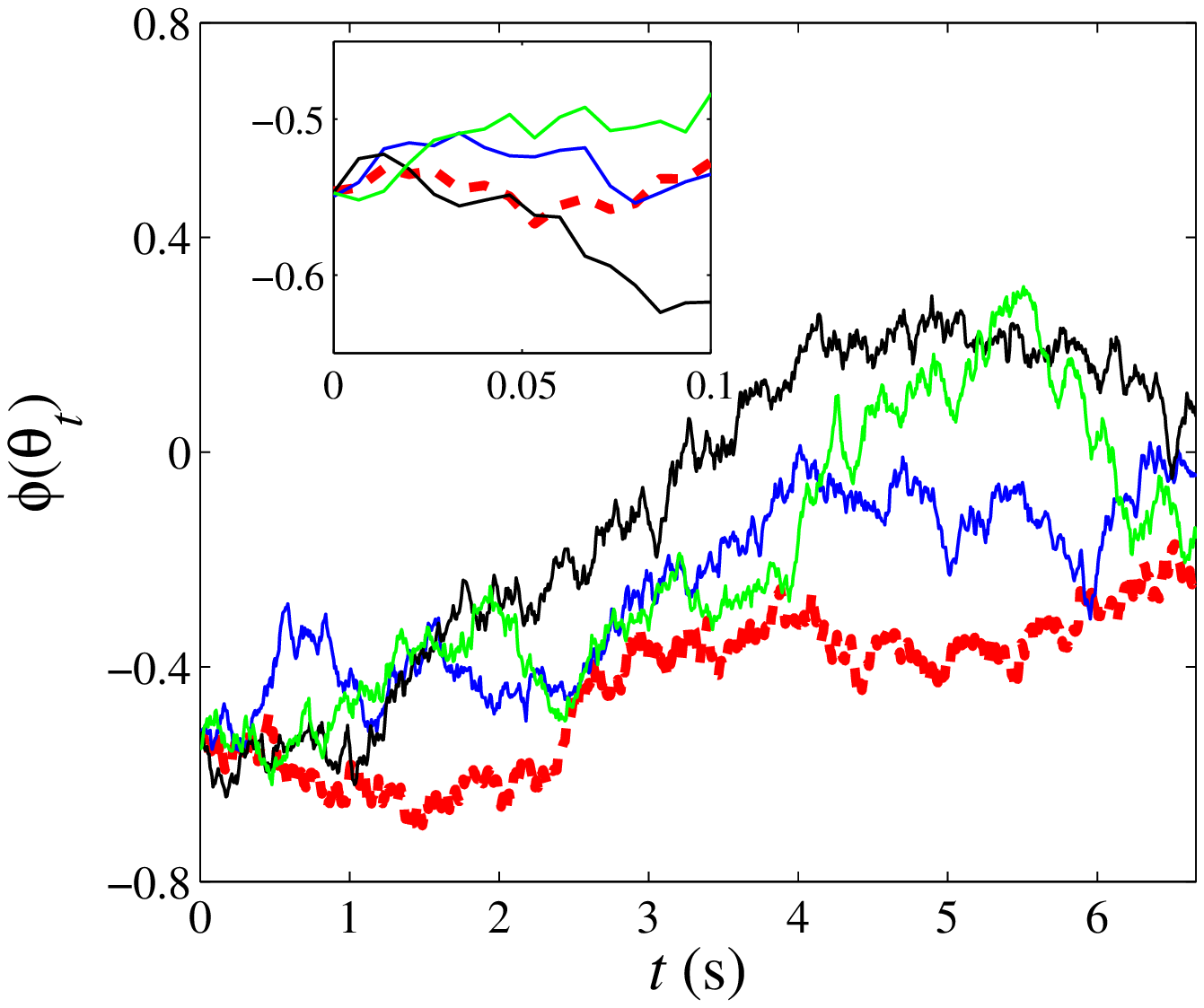}}
     \hspace{.0in}
     \subfigure[]{
          \includegraphics[width=.45\textwidth]{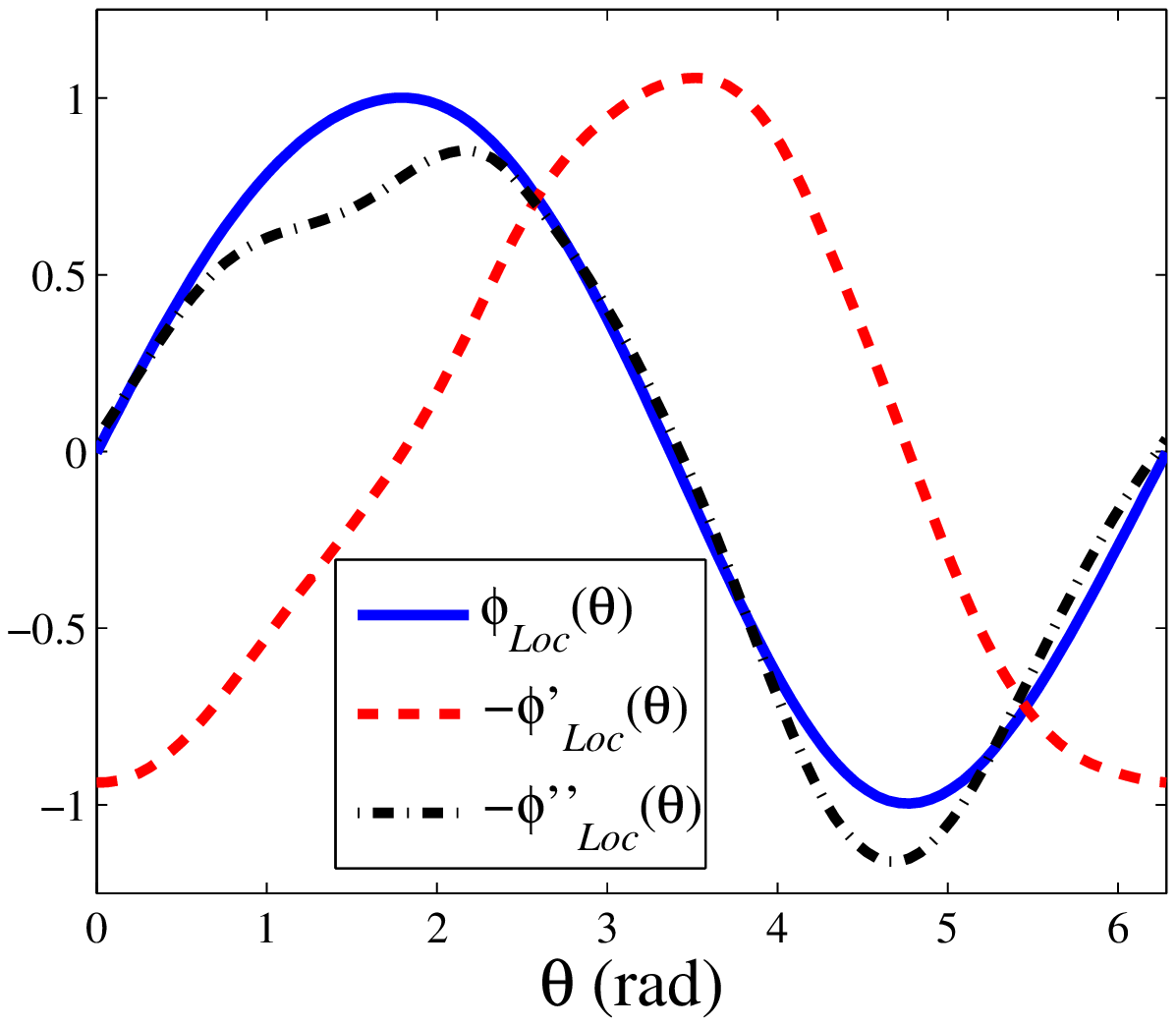}}\\
     \vspace{.0in}
     \subfigure[]{
          \includegraphics[width=.45\textwidth]{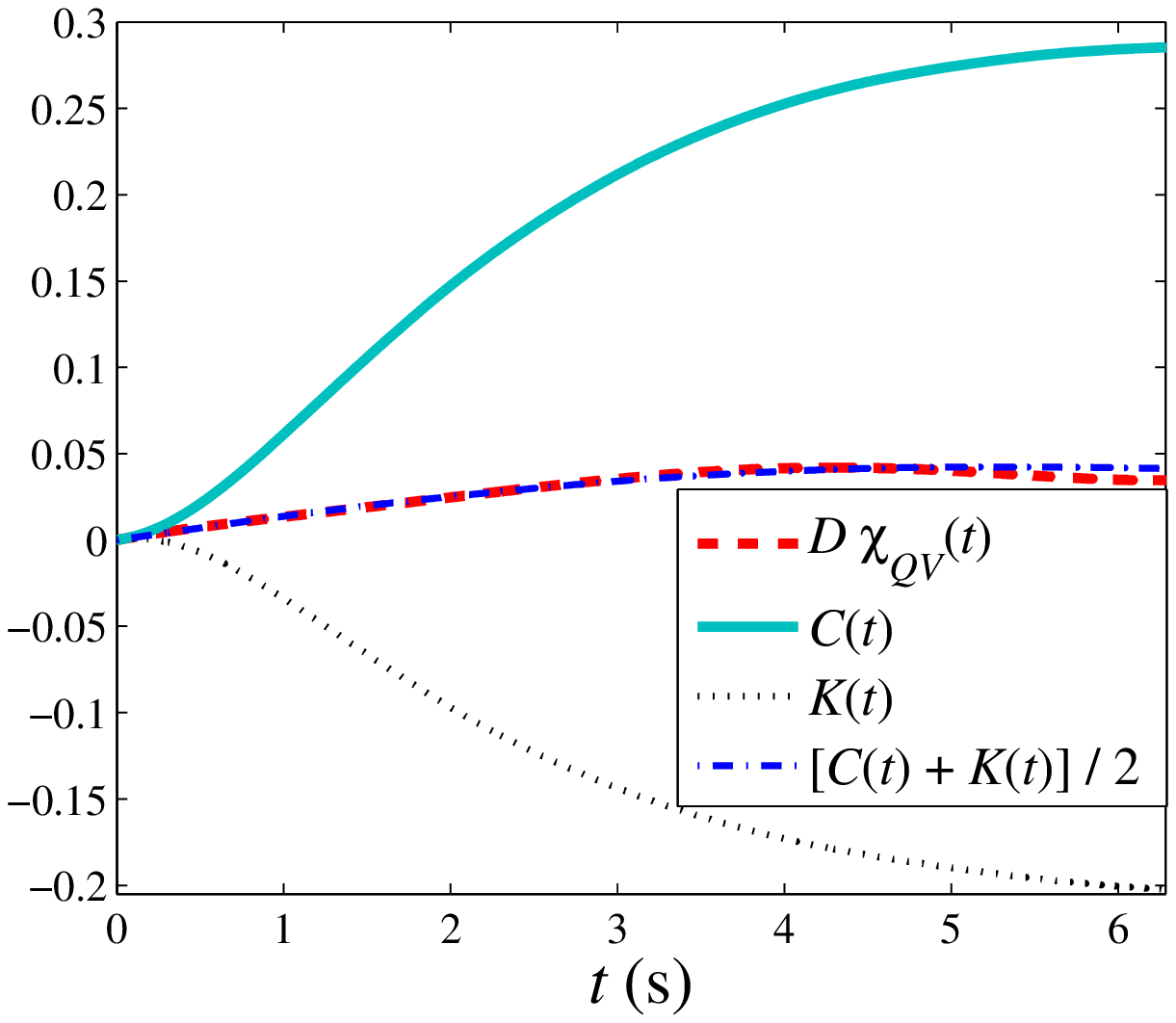}}
     \hspace{.0in}
     \subfigure[]{
          \includegraphics[width=.45\textwidth]{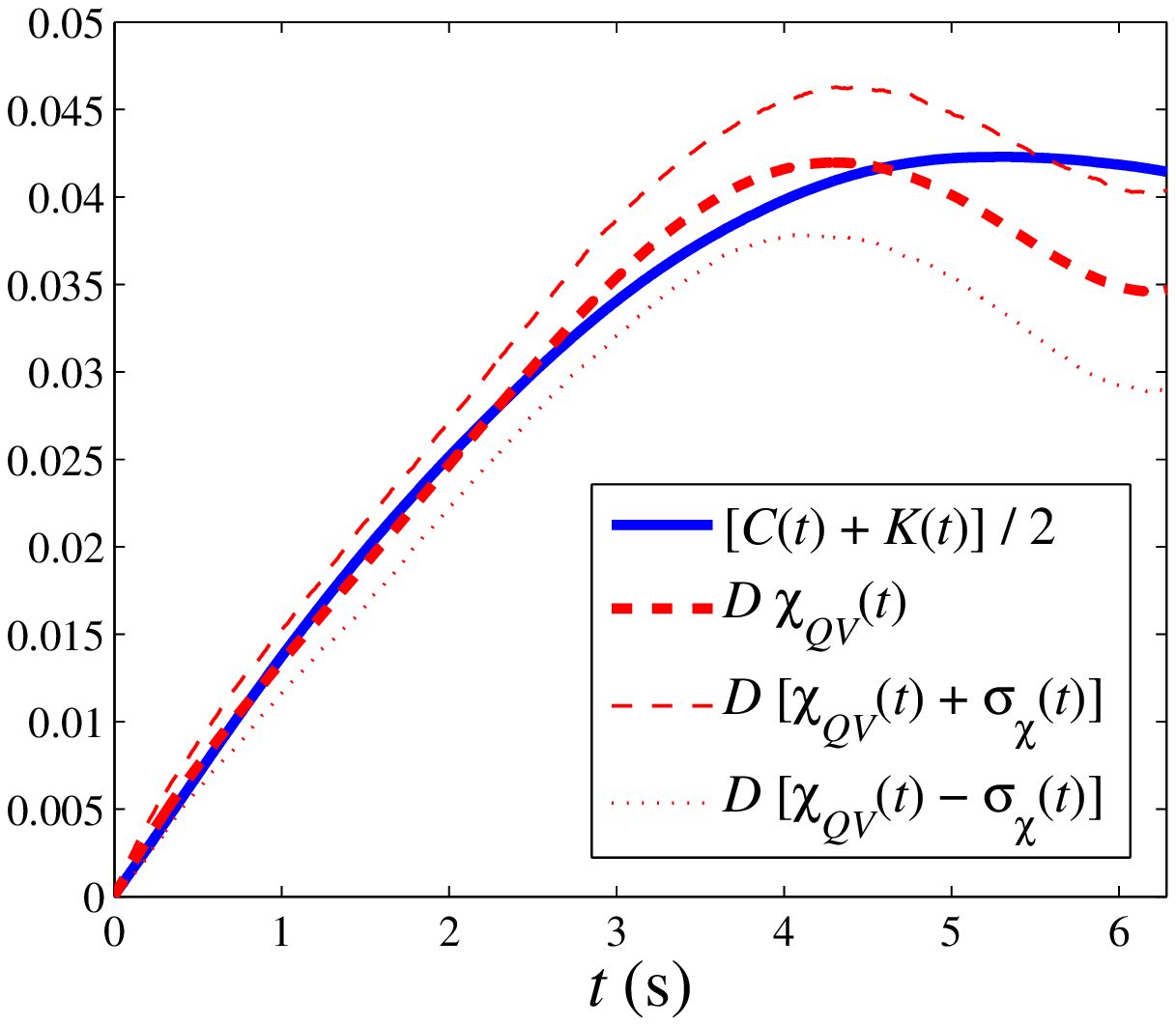}}\\
     \caption{ (a) Examples of a perturbed trajectory (red thick dashed line) and three unperturbed steady-state trajectories (thin solid lines) used to compute the integrated response function given by equation (\ref{eq:intresponse}). Inset: expanded view at short time. The unperturbed trajectories are chosen toç start at the same value as the perturbed one.
 (b) Potential profile $\phi(\theta)$ locally fitted as a third-order polynomial $\phi_{Loc}(\theta)$ around
each value of $\theta$. Their derivatives are computed from $\phi_{Loc}(\theta$). 
(c) Integrated linear response function $\chi_{QV}(t)$, entropic $C(t)$ and frenetic $K(t)$ terms and 
the corresponding indirect measurement of the response $[C(t)+K(t)]/2$ for
$Q=\phi(\theta)$, $h=-\delta A$ and $V=\phi(\theta)$, as functions of the integration time $t$. (d) Expanded view of (c).}
     \label{fig:2}
\end{figure}

\section{Measurement of the linear response function around a non-equilibrium steady state}\label{sec:linear}
Now we proceed to determine the linear response of the particle motion when slightly perturbing the non-equilibrium
steady state previously described. For experimental simplicity we consider a step perturbation to the potential amplitude $A \rightarrow A + \delta A$, so that the perturbation and its conjugate variable are $h = -\delta A$ and
$V(q)=\phi(\theta)$, respectively.
It should be noted that the dynamics of $\theta$ is strongly nonlinear as the particle undergoes the periodic potential $A\phi(\theta)$ but for sufficiently small values of $\delta A$ the response of $\phi(\theta)$ can still be linear. In reference \cite{gomez}, where we used the same experimental data as in the present paper, we have already checked that for a perturbation $\delta A$ of the potential amplitude $A$  in the range $|\delta A| \le 0.07 A$ the linear response regime holds. This was done by computing the experimental integrated response $\chi(t)$ of the observable $Q(\theta)=\sin \theta$ for two different values of $\delta A$: $0.05A$ and $0.07 A$. We showed that within the experimental error bars $\chi(t)$ is the same in both cases. The independence of $\chi(t)$ with respect to $\delta A$ experimentally demonstrates that the system is actually in the linear response regime at least for $|\delta A| \le 0.07 A$ and times $0 \le t \le 3.3$ s after the application of the Heaviside perturbation. 

In order to illustrate the meaning of the non-equilibrium linear
response relation of equation (\ref{eq:intGFD}) in this case, we perform two different kinds of independent
measurements: \emph{direct} and \emph{indirect}, as explained in the following.

\subsection{Direct measurement of the linear response function}\label{subsec:direct}
First, we consider the \emph{direct} measurement of the integrated response function $\chi_{QV}$ for the periodic observable
$Q(q)=\phi(\theta) = \phi(\theta+2\pi)$. This observable times $\gamma a^2 A$ represents the instantaneous potential
energy of the particle. This is experimentally accomplished by applying a step perturbation
to the amplitude of the sinusoidal laser power modulation, 4.2 mWpp $\rightarrow$ 4.4 mWpp, but keeping constant the power offset at 30 mW so that
$F$ remains constant. This results in a perturbation of the potential amplitude $\delta A = 0.05 A$.
Then, the integrated response function
of $\phi(\theta)$ at time $t\ge 0$ due to the perturbation $-\delta A$ applied at time $0$ is given by
\begin{equation}\label{eq:intresponse}
    \chi_{QV}(t) = \frac{\langle \phi(\theta_t) \rangle_{\delta A} - \langle \phi(\theta_t) \rangle_0}{-\delta A},
\end{equation}
where $\langle ... \rangle_{\delta A}$ and $\langle ... \rangle_{0}$  denote the ensemble averages of the
perturbed $\theta_{t,\delta A}$ and unperturbed $\theta_t$ trajectories, respectively.
In order to decrease the statistical errors in comparison of the terms in equation (\ref{eq:intresponse}),
for a given perturbed trajectory $\theta_{t,\delta A}$ we look for as many unperturbed ones
$\theta_t$ as possible starting at time $t^*$ such that $\phi(\theta_{t^*})=\phi(\theta_{0,\delta A})$.
Then we redefine $t^*$ as $t=0$ in equation (\ref{eq:intresponse}), as depicted in figure \ref{fig:2}(a). The unperturbed trajectories found in this way
allow us to define a subensemble over which the average $\langle ... \rangle_{0}$ is computed.
On the other hand, the average $\langle ... \rangle_{\delta A}$ is performed over 500
independent realizations of $\delta A$ that are enough for a fair
determination of the integrated response.
The resulting curve $\chi_{QV}(t) $
and its corresponding error bars $\pm \sigma_{\chi}(t)$ as
functions of the integration time $t$ starting at the instant of
the application of the step perturbation are shown in
figures \ref{fig:2}(c) and \ref{fig:2}(d).

\subsection{Indirect measurement of the linear response function}\label{subsec:indirect}
The same response information can be obtained \emph{indirectly}
from correlation measurements of the unperturbed non-equilibrium steady-state fluctuations ($\delta A=0$) of $\theta_t$
when properly using equation (\ref{eq:intGFD}).
For the Langevin dynamics of $\theta$ described by equation (\ref{eq:1stLangevin}) the analytical expression of the generator $L$ is
\begin{equation}\label{eq:generator1st}
    L = (F-A \phi'(\theta))\partial_{\theta}+D\partial_{\theta}^2.
\end{equation}
Hence, in this case equation (\ref{eq:intGFD}) reads
\begin{equation}\label{eq:intGFD1st}
    D \chi_{QV}(t)=\frac{C(t) + K(t)}{2},
\end{equation}
where the entropic and frenetic terms are
\begin{equation}\label{eq:entropic1st}
    C(t)=\langle \phi(\theta_t) \phi(\theta_t) \rangle_0-\langle \phi(\theta_0) \phi(\theta_t)\rangle_0,
\end{equation}
\begin{equation}\label{eq:frenetic1st}
    K(t)=-\int_0^t ds \langle [D \phi''(\theta_s) + (F-A\phi'(\theta_s)) \phi'(\theta_s)] \phi(\theta_t)\rangle_0,
\end{equation}
respectively. At this point it is clear that we need to know the potential profile 
$\phi(\theta)$ for the
indirect method. The integrand of
equation (\ref{eq:frenetic1st}) involves the instantaneous values of
$\phi(\theta)$ and its derivatives $\phi'(\theta)$ and
$\phi''(\theta)$. In order to take into
account the non-sinusoidal distortion of the potential profile, we perform a local polynomial fit $\phi_{Loc}$ of $\phi$ around
each value of $\theta \in[0,2\pi)$. Then the instantaneous value of the observable
$\phi(\theta_t)$ at time $t$ is approximated by
$\phi_{Loc}(\theta_t)$ either for an unperturbed or a perturbed
trajectory. The local polynomial approximation $\phi_{Loc}$ and its
derivatives $\phi_{Loc}'$, $\phi_{Loc}''$ are plotted in
figure \ref{fig:2}(b) showing the non-sinusoidal distortion. 

The resulting curves $C(t)$ and $K(t)$ as functions of the integration
time $t$ are plotted in figure \ref{fig:2}(c). 
At thermal equilibrium ($F=0$) one should find that $C(t)=K(t)$ for all $t
\ge 0$ because of the time reversibility and stationarity of the
two-time correlations leading to the equilibrium
fluctuation-dissipation relation $D\chi_{QV}(t)=C(t)$. On the other hand,
in the present case $K(t)$ reaches negative values of
the same order of magnitude as the positive values of $C(t)$. This
reflects the experimental conditions far from thermal equilibrium
of the system. 
The curve for $K(t)$ represents the
first experimental result concerning the direct measurement of the dynamical activity along a trajectory \cite{baiesi1,baiesi2,baiesi3}.
The average of these two quantities $[C(t)+K(t)]/2$
is one order of magnitude smaller. This average, which is an
indirect measurement of the integrated response function according
to equation (\ref{eq:intGFD1st}), agrees very well with the direct
measurement of $\chi_{QV}$ within the experimental error bars, as
shown in figure \ref{fig:2}(d).
All is consistent with the results of \cite{gomez}, see \cite{chetrite,baiesi1}, but the experimental approach here is quite different.
The experimental verification of previous equivalent generalized fluctuation-dissipation relations for the same experimental system as the one studied here \cite{gomez,blickle1,mehl} involves the local mean velocity of the particle (\emph{i.e.} the ratio between the steady-state current and the probability density). Here we measure directly explicit correlation functions $C(t)$ and $K(t)$ without recourse to and indeed without need for the expression for the stationary distribution. Note that the experimental density profile $\rho(\theta)$ (and not its analytical expression) is used in practice only for calibration purposes \cite{gomez,blickle} to determine the experimental parameters $A$ and $F$. However,  once these parameters are known the data analysis necessary to compute $C(t)$ and $K(t)$ completely relies on the dynamics, i.e. the measurement of the time series $\theta_t$. Thus, this procedure is suitable for the study of the linear response of more complex Markovian non-equilibrium situations  ({\emph{e.g.} multiple degrees of freedom, non-stationary states, several thermal baths) where local mean velocities are not easily measurable in experiments.

\subsection{Discussion}
We remark that for this kind of micron-sized system
the relation  (\ref{eq:intGFD}) actually represents
a feasible indirect method to access the linear response regime
far from thermal equilibrium. This is because all the parameters
of the unperturbed dynamics are known a priori or can be
determined in situ without any external perturbation of the
non-equilibrium state. On the other hand, the direct measurement of the
linear response function exhibits a number of technical
difficulties in practice. First, a vanishingly small Heaviside perturbation
$-h V(q)$ to the initially unperturbed potential $U(q)$
is ideally required. Otherwise spurious
effects quickly bias the measurement of $\chi_{QV}$, specially
when the system is strongly non-linear. 
Second, one requires an extremely large number of independent
realizations of $h$ to resolve $\chi_{QV}$ as the perturbation
$-hV(q)$ must be chosen very weak, typically smaller than the
thermal fluctuations of the energy injected by the environment.
This is evident on the results of the direct
measurement of $\chi_{QV}$ for the colloidal particle, see
figure \ref{fig:2}(d). For integration times $t \lesssim 3$ s the
agreement between $D\chi_{QV}(t)$ and $[C(t)+K(t)]/2$ is
excellent. Then, for $t \gtrsim 3$ s
the finite sampling leads to deviations between the two methods and
increasingly large error bars for the direct measurement of
$\chi_{QV}$. These drawbacks are skipped when implementing the
indirect method measuring the unperturbed quantity
$[C(t)+K(t)]/2$. In addition, for a steady state like the one
experimentally studied here one can improve dramatically the
statistics by performing an additional time average over a window
$[0,t_{max}]$
\footnote{The only restriction in the choice of $t_{max}$ is that the ensemble of 
averaging intervals $[\theta_0,\theta_{t_{max}}]$ must cover the whole 
circle to sample correctly the steady state.}
: $\bar{C}(t)= \int_{0}^{t_{max}}
C(t+u)\mathrm{d}u/t_{max}$, $\bar{K}(t)= \int_{0}^{t_{max}}
K(t+u)\mathrm{d}u/t_{max}$, as actually done for the
curves $C$, $K$ and $(C+K)/2$ in figures \ref{fig:2}(c) and
\ref{fig:2}(d). However, one must be careful when performing the
time average. This is because the value of $t_{max}$ may affect the resulting values of
$[C(t)+K(t)]/2$ as $t$ increases, specially for correlation
functions involving strongly fluctuating quantities such as
velocities, as recently discussed in \cite{mehl}. For the curves
shown in figures \ref{fig:2}(c) and \ref{fig:2}(d) we verified that
their shapes are not significantly influenced by $t_{max}$.

\begin{figure}
     \centering
     \subfigure[]{
          \includegraphics[width=.45\textwidth]{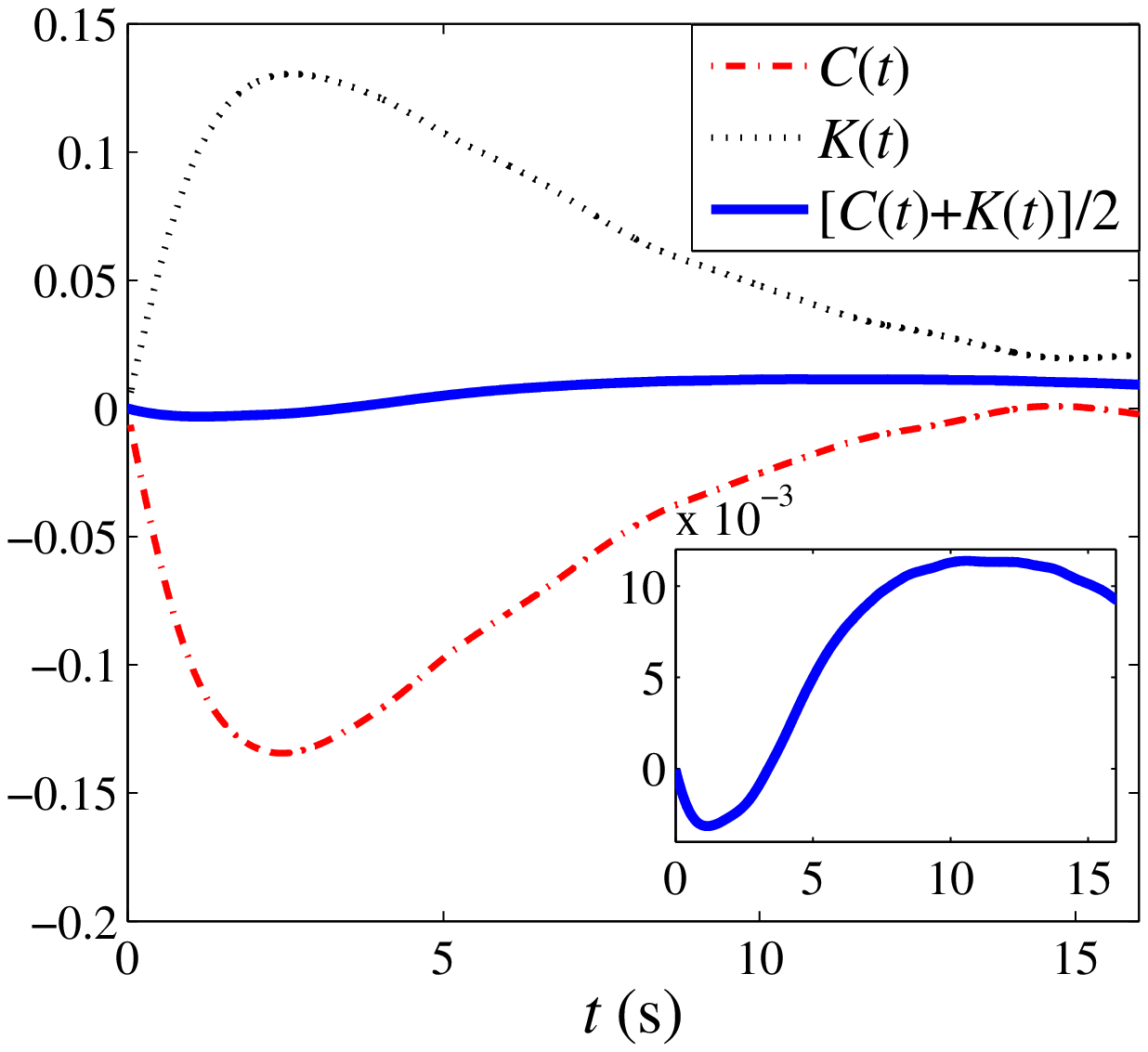}}
     \hspace{.0in}
     \subfigure[]{
          \includegraphics[width=.45\textwidth]{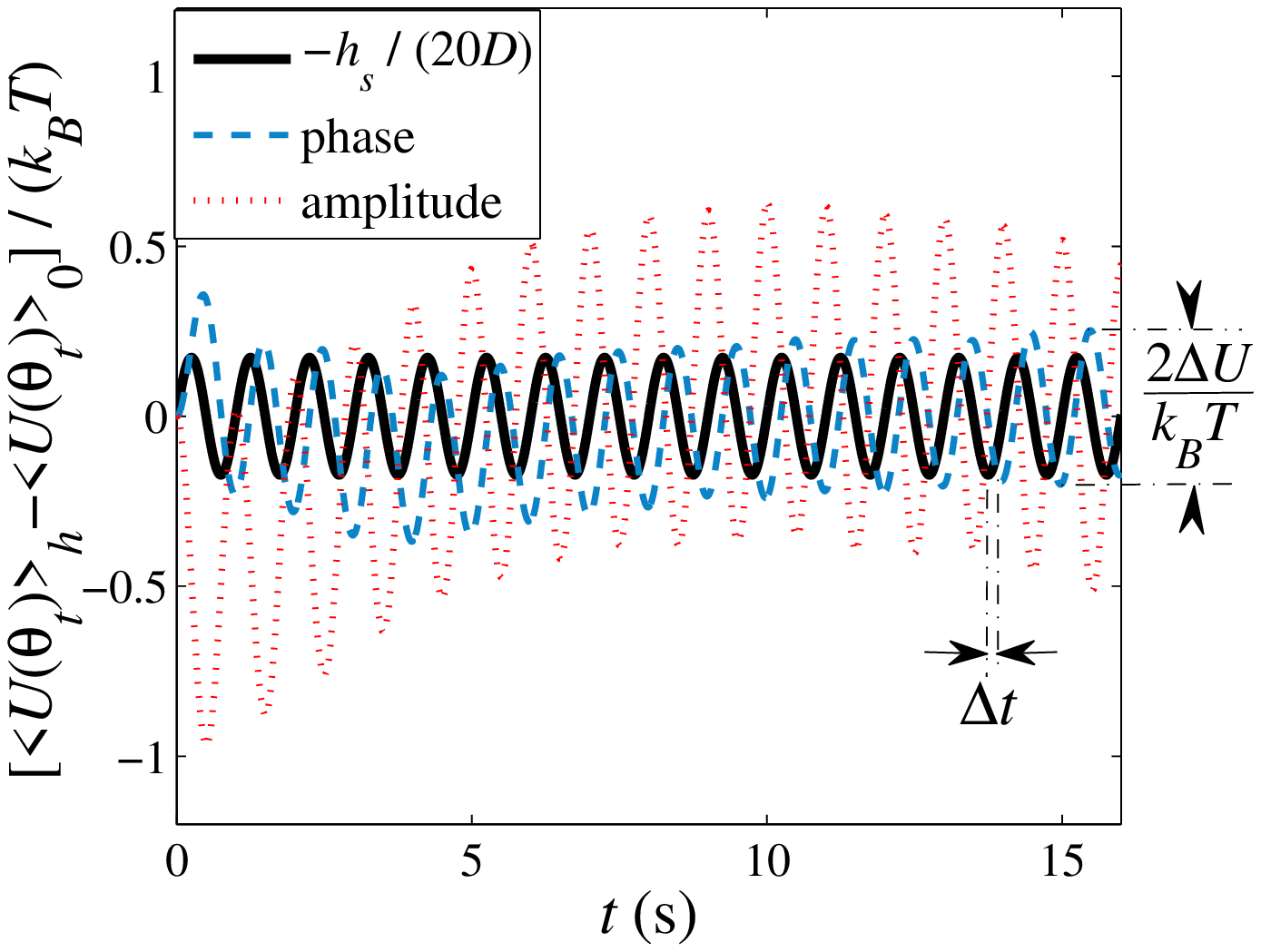}}\\
     \vspace{.0in}
     \subfigure[]{
          \includegraphics[width=.45\textwidth]{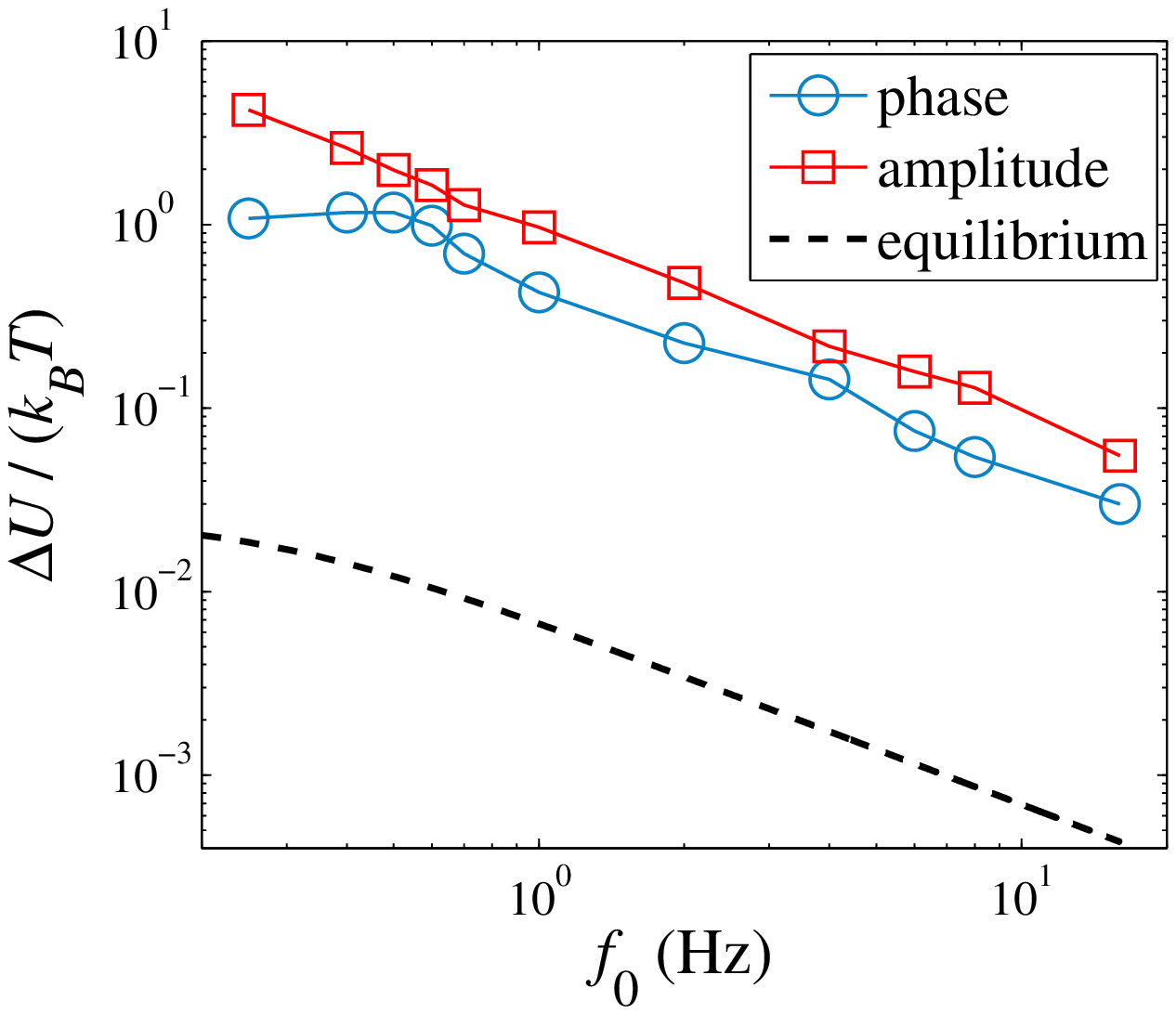}}
     \hspace{.0in}
     \subfigure[]{
          \includegraphics[width=.45\textwidth]{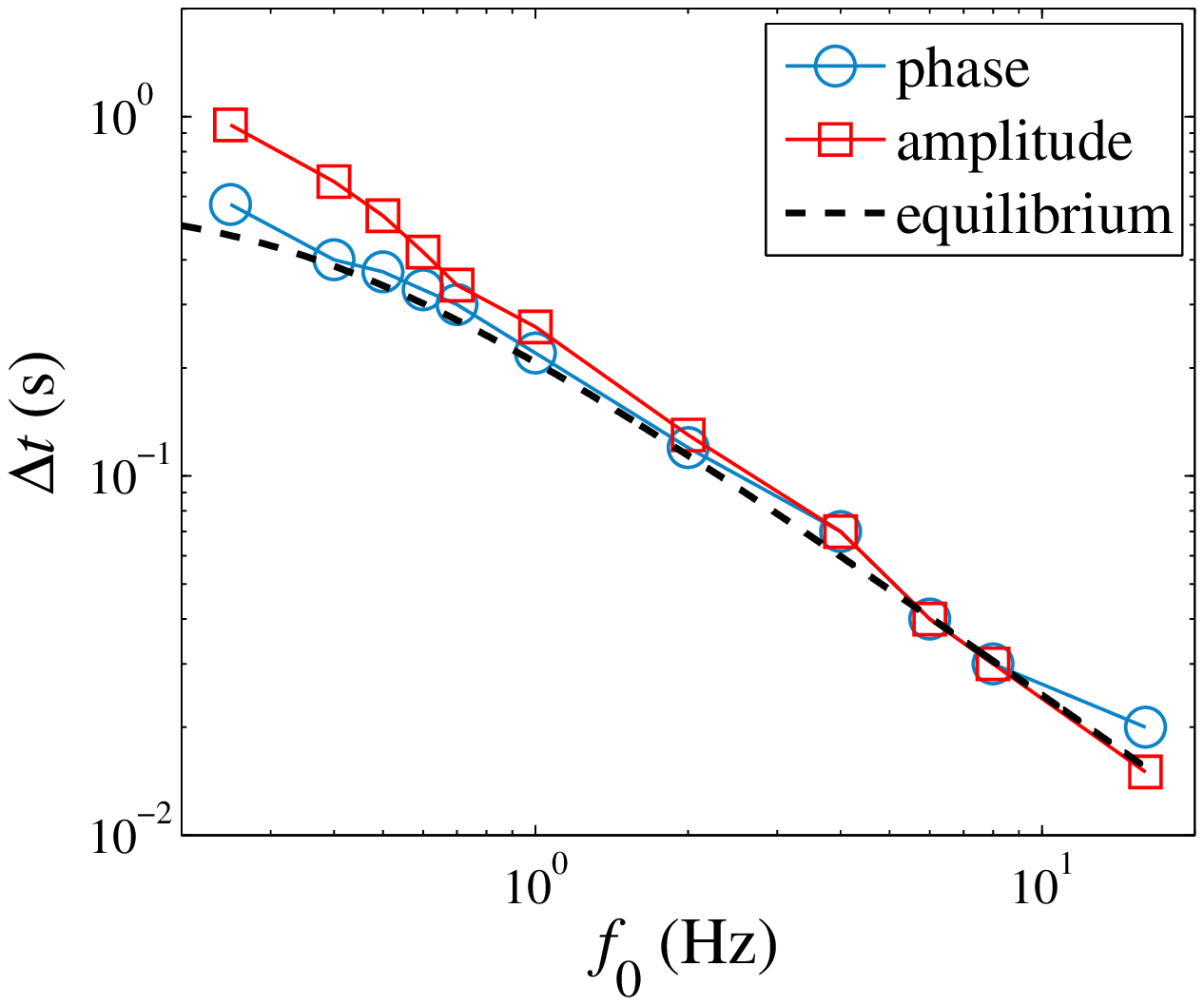}}\\
     \caption{ a) Integrated response function of the observable $Q=\phi(\theta)$ for a small perturbation of the potential
phase as a function of the integration time $t$.
Inset: expanded view of $[C(t)+K(t)]/2$. (b) Sinusoidal time-dependent
perturbation $-h_s$ (solid black line) of the static potential $A\phi$.
Resulting mean potential energy of the Brownian particle for a phase perturbation (dashed blue line) 
and an amplitude perturbation (dotted red line) for $-h_0/A=0.05$ and $f_0=1$ Hz.
(c) Asymptotic values of oscillation amplitude of the potential energy and
(d) the delay time with respect to $-h_s$ for each kind of perturbation.
The black dashed lines represent the values that would be obtained around
thermal equilibrium ($F=0$), given by equations (\ref{eq:amplitude}) and (\ref{eq:delay}).}
     \label{fig:3}
\end{figure}

\subsection{Application example}
Finally, we illustrate the usefulness of the indirect measurement of the linear response function to study the temporal behavior
of the mean potential energy of the particle $\langle U(\theta_t)\rangle_h$ under small external perturbations $h_s$ more intricate than a simple Heaviside function. This is done with the same unperturbed experimental data used in subsection \ref{subsec:indirect} without carrying out the different physical realizations of $h_s$.
We concentrate on a sinusoidal perturbation starting at time $s=0$
\begin{equation}\label{eq:pert} 
    h_s = h_0 \sin 2\pi f_0 s.
\end{equation}
either to the phase or to the amplitude of the potential around the steady state.
First, we consider the case of a small phase perturbation $\alpha_s = \alpha_0 \sin 2\pi f_0 s$ with $\alpha_0\ll 1$:
$A\phi(\theta) \rightarrow A\phi(\theta+\alpha_s)\approx A\phi(\theta) + A\phi'(\theta)\alpha_s$ so that
$h_0=-A\alpha_0$ in equation (\ref{eq:pert}) and $V=\phi'(\theta)$. One must compute first the integrated response function
of $Q=\phi(\theta)$ given by $[C(t)+K(t)]/(2D)$ by inserting the right $V$ and $Q$ in equations (\ref{eq:entropic})
and (\ref{eq:frenetic}). The resulting  curves $C(t)$, $K(t)$ and $[C(t)+K(t)]/(2D)$ are plotted in figure \ref{fig:3}(a).
Next, using equation (\ref{eq:linearresponse})
the experimental impulse response function $[\partial_tC(t-s)+\partial_tK(t-s)]/(2D)$ must be convolved
with $h_s$ given by equation (\ref{eq:pert}). In this way one finds that the mean potential energy of the particle
oscillates around the non-equilibrium steady state value $\langle U(\theta) \rangle_0=-4.7k_B T$
as shown by the dashed blue line in figure \ref{fig:3}(b) for $\alpha_0=0.05$ rad and $f_0=1$ Hz.
The oscillations exhibit a delay time ($\Delta t \approx 0.23$ s) with respect
to $\alpha_s$ and a slow transient ($\sim 15$ s) corresponding to the decay of the non-equilibrium stationary
correlations. As $t$ increases the oscillations settle around $\langle U(\theta) \rangle_0$
with a constant amplitude $\Delta U \approx 0.2k_BT$.
Now, we consider a sinusoidal time-dependent perturbation to the potential amplitude: $\delta A_s = \delta A_0 \sin 2\pi f_0 s$
with the same strength ($-h_0/A = 0.05$) and frequency as before. In this case
$h_0=-\delta A_0$ and $V=\phi(\theta)$. Following the same procedure with $[C+K]/(2D)$ shown in figure \ref{fig:2}(d), one
finds a different qualitative behavior of $\langle U(\theta_t) \rangle_h$, as depicted by the dotted red line in figure \ref{fig:3}(b).
At the beginning the mean potential energy responds in the opposite direction to $\delta A_s$.
Then, as $t$ becomes larger than the slow non-equilibrium transient $\langle U(\theta) \rangle_h$ oscillates
around $\langle U(\theta) \rangle_0$ with a constant amplitude $\Delta U \approx 0.5 k_BT$ and a delay time
$\Delta t \approx 0.26$ s.
For both types of perturbations one can write the asymptotic dependence of  $\langle U(\theta_t) \rangle_h$ 
on $t \gtrsim 15$ s as
\begin{equation}\label{eq:potenergy}
    \langle U(\theta_t) \rangle_h = \langle U(\theta_t) \rangle_0 \pm \Delta U \sin[ 2\pi f_0 (t-\Delta t)],
\end{equation}
where the positive and negative signs stand for the phase and
amplitude perturbations, respectively. The values of $\Delta U$ and
$\Delta t$ depend on the frequency $f_0$. In
figures \ref{fig:3}(c) and \ref{fig:3}(d) we show this dependence.
We now compare these far-from-equilibrium results with those that
would be obtained when applying $h_s$ around thermal equilibrium
($F=0$). In such a case the particle motion is tightly confined to
the harmonic part of the potential around the minimum $\theta_m = 3\pi/2$: 
$\phi(\theta) \approx -1 + (\theta-\theta_m)^2/2$.
After some algebra using this approximation one finds the expression for 
$\langle U(\theta_t) \rangle_h $ when perturbing thermal equilibrium 
\begin{eqnarray}\label{eq:eqpotenergy}
    \langle U(\theta_t) \rangle_h &=& \langle U(\theta_t) \rangle_0  
	\pm \Delta U \{\sin[ 2\pi f_0 (t-\Delta t)] + \nonumber \\ 
      & & e^{-2At} \sin 2\pi f_0 \Delta t \},
\end{eqnarray}
where $\langle U(\theta_t) \rangle_0=-68.3k_BT$ and
\begin{equation}\label{eq:amplitude}
    \Delta U = -\frac{h_0}{A}\frac{k_B T}{2 (1+\pi^2 f_0^2/A^2)^{1/2}} ,
\end{equation}
\begin{equation}\label{eq:delay}
    \Delta t= \frac{1}{2\pi f_0} \arctan\left( \frac{\pi f_0}{A}\right) ,
\end{equation}
either for a phase (positive sign) or an amplitude (negative sign) perturbation. 
Note that for $t\gg (2A)^{-1}$, equation (\ref{eq:eqpotenergy}) exhibits the same 
qualitative behavior as (\ref{eq:potenergy}). We plot the
curves given by eqsuations (\ref{eq:amplitude}) and (\ref{eq:delay}) in
figures \ref{fig:3}(c) and \ref{fig:3}(d), respectively, for the
same values of the parameters $h_0$ and $A$ as before. Unlike the behavior 
close to equilibrium, the oscillation amplitude $\Delta U$
strongly depends on the perturbed parameter around the
non-equilibrum steady state: it is more sensitive to amplitude
perturbations than to phase perturbations. In addition, the
far-from-equilibrium values are two orders of magnitude larger
than that given by equation (\ref{eq:amplitude}). By contrast, the
delay time $\Delta t$ is not significantly affected by the
far-from-equilibrium nature of the system. It is almost independent of
$F$ and of the type of perturbation and it converges to
equation (\ref{eq:delay}) as $f_0$ increases.

\section{Concluding remarks}
We have experimentally studied the non-equilibrium linear response of the potential energy of a Brownian particle
in a toroidal optical trap.
We gain insight into the application of the fluctuation-dissipation relations far from thermal
equilibrium in this non-linear system with a single relevant
degree of freedom. In particular, we show that the entropic-frenetic approach
is appropriate and feasible for the study of the linear response of non-equilibrium micron-sized systems with a small number of degrees of freedom immersed in
simple fluids. Non-trivial linear response information can be obtained
from purely unperturbed measurements of the non-equilibrium fluctuations provided that the parameters
describing the dynamics are known. Our experiment reveals that the indirect
determination of the linear response function is less time-consuming, more accurate and more flexible
than the direct perturbation of the non-equilibrium system.
Similar ideas are expected to be applicable to more complex micron-sized systems such as
atomic-force microscopy experiments and ensembles of colloidal particles in simple non-equilibrium conditions where local mean velocities are difficult to measure.

\end{document}